\documentclass[
   ,final            
  ]
  {aipproc}
\usepackage{natbib}
\layoutstyle{6x9}

\newcommand\ion[2]{#1$\;${\small{#2}}\relax}%
\newcommand {\HI}     {\ion{H}{I}}   
\newcommand {\OVI}    {\ion{O}{VI}}   

\newcommand {\CIII}   {\ion{C}{III}}   

\newcommand {\SIV}    {\ion{S}{IV}}
\newcommand {\CIV}    {\ion{C}{IV}}
\newcommand {\SiIV}   {\ion{Si}{IV}}
\newcommand {\SiIII}  {\ion{Si}{III}}
\newcommand {\FeIII}  {\ion{Fe}{III}}

\newcommand {\kms}    {km~s$^{-1}$}

\newcommand {\etal}  {et~al.}

\begin{document}

\title{Highly Ionized Envelopes of High Velocity Clouds}

\classification{98.62.Ra}
\keywords      {Galactic gas; High velocity clouds}

\author{Erin E. Zekis}{address={University of Colorado, CASA, 389-UCB, Boulder, CO 80309}}
\author{J. Michael Shull}{address={University of Colorado, CASA, 389-UCB, Boulder, CO 80309}}

\begin{abstract}

We present recent results on highly ionized gas in Galactic High-Velocity Clouds (HVCs),
originally surveyed in \OVI~\citep{Sembach2003}. In a new \emph{FUSE}/HST survey of \ion{Si}{II/III/IV}
(Shull \etal~2009) toward 37 AGN, we detected \SiIII~($\lambda$1206.500 \AA)
absorption with a sky coverage fraction $81 \pm$ 5\% (61 HVCs along 30 of 37 high-latitude
sight lines). The \SiIII~($\lambda$1206.500 \AA) line is typically 4-5
times stronger than \OVI~($\lambda$1031.926 \AA).
The mean HVC colum density of perhaps $10^{19} {\rm cm}^{-2}$ of low-metallicity ($0.1-0.2 Z_{\odot}$)
ionized gas in the low halo. Recent determinations of HVC distances allow us to estimate
a total reservoir of $\sim 10^8 M_{\odot}$. Estimates of infall velocities indicate an
infall rate of around $1 M_{\odot} {\rm yr}^{-1}$, comparable to the replenishment rate for star
formation in the disk. HVCs appear to be sheathed by intermediate-temperature gas ($10^{4.0}-10^{4.5} K$)
detectable in \SiIII~and \SiIV, as well as hotter gas seen in \OVI~and other high ions. To prepare for
HST observations of 10 HVC-selected sight lines with the {\it Cosmic Origins Spectrograph} (COS),
we compile \emph{FUSE}/STIS spectra of these ions, plus \FeIII, \CIII, \CIV, and \SIV. Better
constraints on the physical properties of HVC envelopes and careful treatment of HVC kinematics
and infall rates should come from high-quality (S/N $\approx$ 30-40) COS data.

\end{abstract}

\maketitle


\section{A Brief History of HVCs}

High Velocity Clouds (HVCs) were first detected in \HI~21 cm emission by \cite{Muller}; quasar sight lines
showed emission at high velocity. Until
the last few years, studies of these objects were unable to determine their distances, and therefore masses.
There has been much debate about their composition, origin, and other basic features.
HVCs are characterized by anomalous velocities with respect to Galactic rotation, with $|v_{LSR}| >$ 90 \kms.
Since 1963, HVCs have been detected in \OVI, \CIV, \SiIV, and various low ions
\citep[see][]{Sembach2003,CSG03,CSG07,Fox04}. The presence of both low and high ions in HVC sight lines may
suggest shock ionization as gas falls into the disk \citep{CSG07,Fox04}.

\section{Recent Progress}

\begin{figure}[t]
  \includegraphics[height=0.55\textheight]{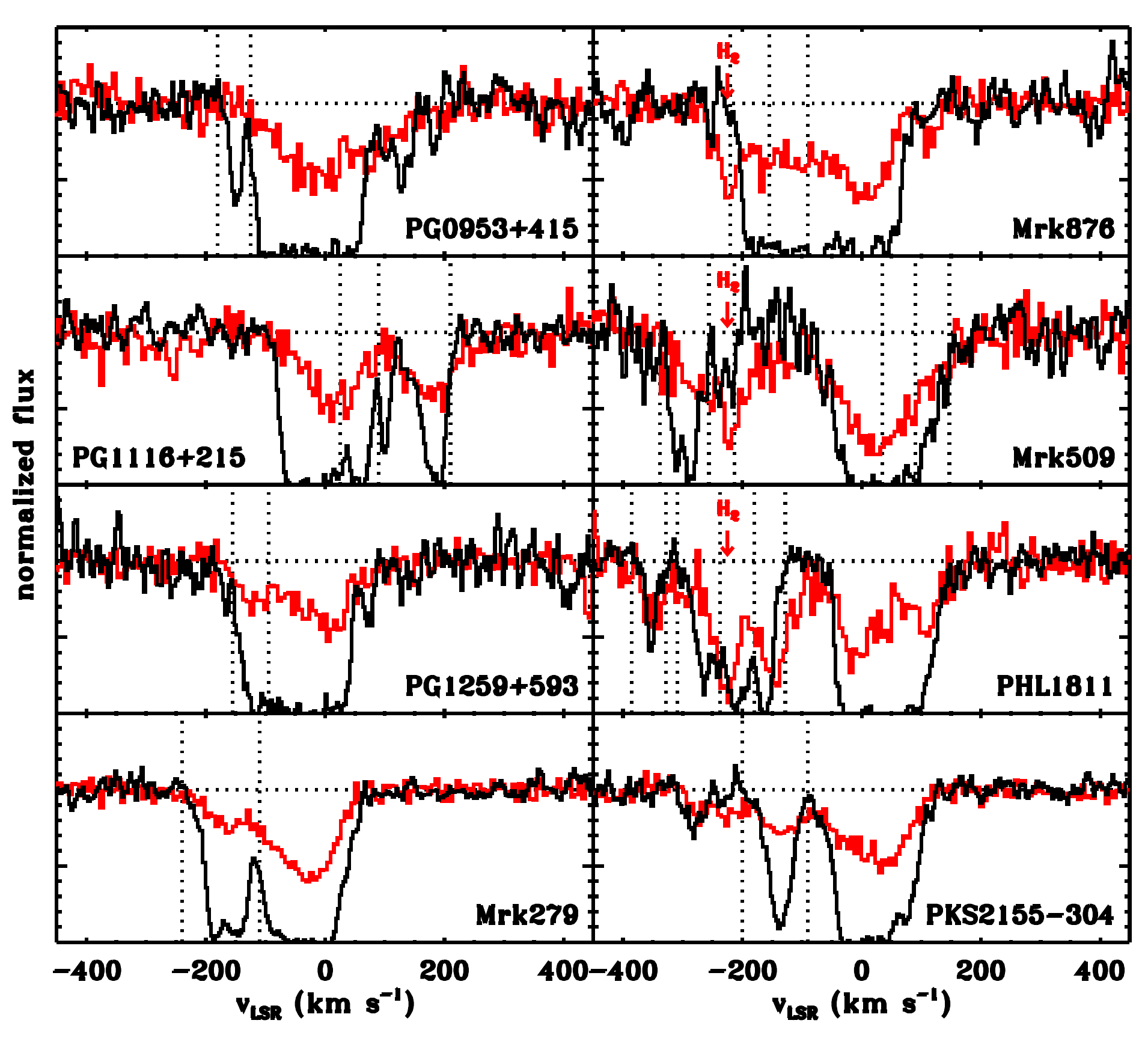}
  \caption{\label{fig1} Comparison plot of several sight lines from Shull \etal~(2009) \citep{Shull2009} displaying the difference between the sensitivity of \OVI~($\lambda$1032, red/gray) and \SiIII~($\lambda$1206, black). Interstellar H$_2$ lines are marked.
  }
\end{figure}

Shull \etal~(2009) \citep{Shull2009} find \SiIII~to be a highly sensitive detector
of HVCs (Figure \ref{fig1}).
Along several of the 37 sight lines in their study, they find
several small cloud components, partly because \SiIII~$\lambda1206$ is typically 4-5 times stronger than
\OVI~$\lambda1032$. The mean column density in their sample is <log $N_{Si~III}$> = 13.44 $\pm$ 0.21. Using
the photoionization code Cloudy, they constrain the mean photoionization parameter to be <log $U$> = -3.0 $\pm$ 0.2.
This suggests mingling of low and high ions due to shock-ionization, as gas
falls through the Galactic halo, with a low ion cloud core shielded inside a hotter envelope.

The distance to HVC Complex C, an extensive cluster of sight line detections covering $\sim$ 1600
square degrees on the sky,
has been determined by \cite{Thom08} to be 10 $\pm$ 2.5 kpc. \cite{Westmeier}
detect HVCs at similar altitudes around M31. If this is a characteristic distance for most
HVCs, this means they constitute a total mass reservoir of $\sim 10^8 M_{\odot}$ and have an infall rate
on the order of $1 M_{\odot} yr^{-1}$. HVC infall may be sufficient to
replenish star-forming gas in the Galactic disk. The source of the gas is still
not well constrained; however, \cite{CSG07} have found that HVCs have low metallicities, $\approx$ 0.1-0.2
$Z_{\odot}$, which points to mixing of primordial and Galactic gas.

\section{Current and Future Data: Detection in High Ions}

The goal of our research will be to use accurate kinematics and high S/N
spectroscopy to determine mass infall rate and ionization states of the HVCs.
The \emph{Hubble Cosmic Origins Spectrograph} (COS) will observe 10 sight lines with \OVI -selected HVCs.
Over the course of 28 orbits, we will obtain spectra with S/N $\approx$ 30-40.
In preparation, we have compiled data from \emph{FUSE} and \emph{Hubble}
STIS. All 10 sight lines have \OVI~data from \emph{FUSE}, but only a few have STIS data. The
sight lines selected include some well-studied targets, such as Mrk 876, as well as some sight lines that
currently have only low-S/N data, such as ESO 265-G23.


\begin{theacknowledgments}
Thanks to Charles Danforth and Joseph Collins for help with data analysis and plotting techniques.
Research and travel funding for this work come from COS grant NNX08-AC14G at the University of Colorado.
\end{theacknowledgments}

\bibliographystyle{aipproc}   

\bibliography{zekis_erin_confproc}

\end{document}